# 3D Etching Profile Evolution Simulation Using Sparse Field Level Set Method

B. Rađenović, S.J. Kim, and J. K. Lee [a]

[a] *Department of Electronic and Electrical Engineering, Pohang University of Science and Technology, Pohang 790-784, South Korea*

Level set method, introduced by Osher and Sethian [1], is a highly robust and accurate computational technique for tracking of moving interfaces in etching, deposition and photolithography processes. It originates from the idea to view the moving front as a particular level set of a higher dimensional function, so the topological merging and breaking, sharp gradients and cusps can form naturally, and the effects of curvature can be easily incorporated. The corresponding equations of motion for the propagating surfaces, which resemble Hamilton-Jacobi equations with parabolic right-hand sides, can be solved using methods for solving hyperbolic conservation laws, ensuring in that the way correct entropy-satisfying solution [2]. In this paper we describe an application of the sparse field method for solving level set equations in 3D plasma etching simulations. Sparse field method itself, developed by Whitaker [3] and broadly used in image processing community, is an alternative to the usual combination of narrow band and fast marching procedures for the computationally effective solving of level set equations. The obtained results for convex and non-convex Hamiltonians show correct behavior of etching profiles in simple model cases.

## SPARSE-FIELD LEVEL SET METHOD

The basic idea behind level set method is to represent surface in question at a certain time $t$ as the zero level set (with respect to the space variables) of a certain function $\varphi(t, \mathbf{x})$, the so called level set function. The initial surface is given by $\{\mathbf{x} \mid \varphi(0, \mathbf{x}) = 0\}$. The evolution of the surface in time is caused by "forces" or fluxes of particles reaching the surface in the case of the etching process. The velocity of the point on the surface normal to the surface will be denoted by $F(t, \mathbf{x})$, and is called velocity function. For the points on the surface this function is determined by physical models of the ongoing processes; in the case of etching by the fluxes of incident particles and subsequent surface reactions. The velocity function generally depends on the time and space variables and we assume that it is defined on the whole simulation domain.

At a later time $t > 0$, the surface is as well the zero level set of the function $\varphi(t, \mathbf{x})$, namely it can be defined as a set of points $\{\mathbf{x} \in \Re^n \mid \varphi(t, \mathbf{x}) = 0\}$. This leads to the level set equation

$$\frac{\partial \varphi}{\partial t} + F(t, \mathbf{x}) \|\nabla \varphi\| = 0 , \qquad (1)$$

in the unknown function $\varphi(t, \mathbf{x})$, where $\varphi(0, \mathbf{x}) = 0$ determines the initial surface. Having solved this equation the zero level set of the solution is the sought surface at all later times. Actually, this equation relates the time change to the gradient via the velocity function. In the numerical implementation the level set function is represented by its values on grid nodes, and the current surface must be extracted from this grid. In order to apply the level set method a suitable initial function $\varphi(0, \mathbf{x})$ has to be defined first. The natural choice for the initialization is the signed distance function of a point from the given surface. This function is the common distance function multiplied by -1 or +1 depending on which side of the surface the point lies on.

As already stated, the values of the velocity function are determined by the physical models. In the actual numerical implementation equation (1) is represented by the upwind finite difference schemes (see ref. [4] for the details) that requires the values of this function at the all grid points considered. In reality the physical models determine the velocity function only at the zero level set, so it must be extrapolated suitably at grid points not adjacent to the zero level set.

Several approaches for solving level set equations exist which increase accuracy while decreasing computational effort. They are all based on using some sort of adaptive schemes.

The most important are narrow band level set method [4], widely used in etching process modeling tools (for a detailed review see [5]), and recently developed sparse-filed method [2], implemented in medical image processing ITK library [6]. Adaptive methods use the fact that actual calculations should not be performed for points far away from the zero level set, since these points do not have any influence. This is the starting assumption in narrow band methods; the width of the narrow band is predefined and should be as small as possible. In actual implementations it is necessary to choose a new narrow band whenever the front hits the boundary of the current narrow band. Another problem is to find the balance between the width of the narrow band and the frequency of reinitializations. This technique provides a substantial speed up; in three dimensions the computational effort is reduced from $O(n^3)$ to $O(n^2)$ compared to fixed grids on fixed simulation domains.

The sparse-field method use an approximation to the distance function that makes it feasible to recompute the neighborhood of the zero level set at each time step. In that way, it takes the narrow band strategy to the extreme. It computes updates on a band of grid points that is only one point wide. The width of the neighborhood is such that derivatives for the next time step can be calculated. This approach has several advantages. The algorithm does precisely the number of calculation needed to compute the next position of the zero level set surface. The number of points being computed is so small that it is feasible to use a linked-list to keep a track of them, so at each iteration only those points are visited whose values control the position of the zero level set surface. As a result, the number of computations increases with the size of the surface, rather than with the resolution of the grid. In fact, the algorithm is analogous to a locomotive engine that lays down tracks before it and picks up them up behind.

**SIMPLE ETCHING MODEL RESULTS**

The details about the code design and implemented algorithms will be published elsewhere. Here we will present some results of preliminary calculations. First, we shall present some illustrating figures, in which we didn't try to relate etching rate (velocity function $F(t, \mathbf{x})$) with the etching conditions (particle fluxes and surface processes). Instead, we used simple analytical expressions describing typical model etching rates [5]. All calculations are performed on 128×128×384 rectangular grid. In Figure 1. the initial profile surface is shown, together with the boundaries of the computational domain. Above the profile surface is the trench region; the top quadrangle is the surface from which the particles involved in etching process come from.

The evolution of the etching profile with time is shown in following figures. In Figure 2. the results obtained for constant velocity function are shown. Behavior of the etching profile is as expected, and the rounding of the corners is reproduced correctly.

The more sophisticated set of examples arises in simulations in which etching rate depends on the angle of the incidence of the incoming particles. In the cases under study here we shall consider an etching beam coming down in the vertical direction. These conditions are characteristic for ion milling technology, but angular dependence of the etching rates appears, more or less, in all etching processes. If write level set equations in Hamilton-Jacobi form, this angular dependence sometimes leads to so-called non-convex Hamiltonians. In order to obtain acceptable results using upwind finite difference filters implemented in ITK library [6], we have used in the following calculations level set equation with additional curvature term:

$$\frac{\partial \varphi}{\partial t} + F(t, \mathbf{x}) \|\nabla \varphi\| + \gamma \kappa \|\nabla \varphi\| = 0, \qquad (2)$$

where $\gamma$ is a small constant, and $\kappa$ is the mean curvature given by

$$\kappa = \frac{1}{2} \nabla \cdot \frac{\nabla \varphi}{\|\nabla \varphi\|}. \qquad (3)$$

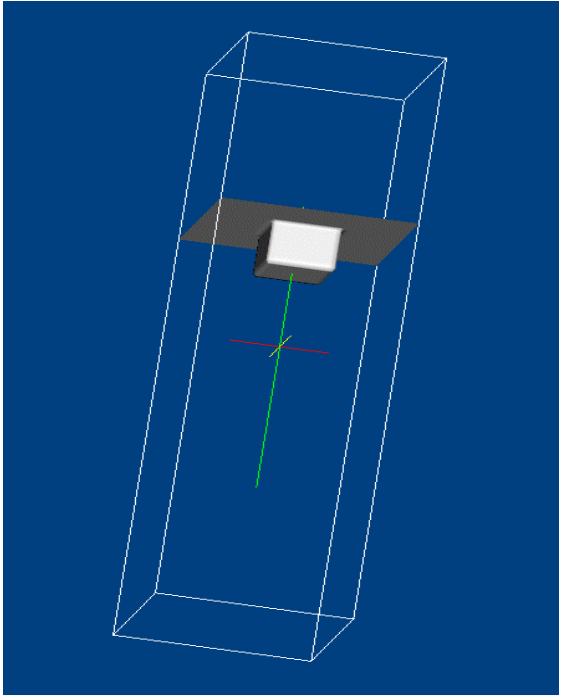

**Figure 1.** Initial profile

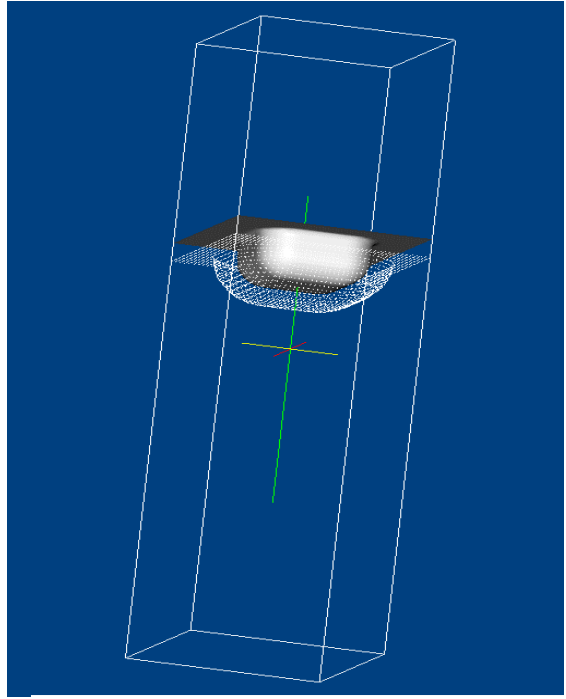

**Figure 2.** Etching profile at t = 5 and 10 (arbitrary units). Velocity function $F(t, \mathbf{x}) = 1$.

In fact, the effects of this dissipative term should be similar to the ones of the Lax-Friedrichs difference scheme, which is a standard choice for this type of calculations.

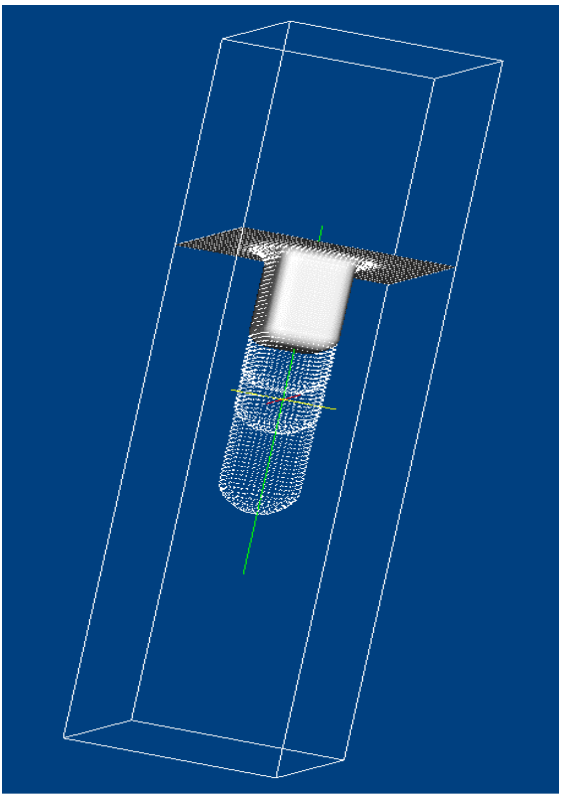

**Figure 2.** Etching profile at t = 20, 40, 60 and 100 (arbitrary units). Velocity function $F(t, \mathbf{x}) = cos(\theta)$.

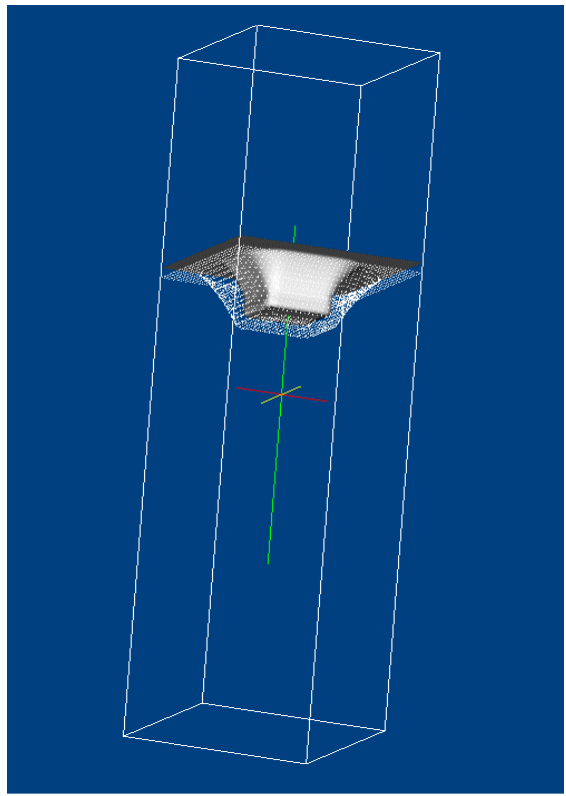

**Figure 1.** Etching profile at t = 5 and 20 (arbitrary units). Velocity function $F(t, \mathbf{x}) = cos(\theta) [1 + 4\ sin^2(\theta)]$.

In Figure 3. the shape of the etching profile in the case when etching rate is proportional to $cos(\theta)$, where $\theta$ is the angle between the surface normal and the route of the incoming particles. This is simplest form of angular dependence, but often describes etching process correctly.

As it can be seen from the figure, the horizontal surfaces move downward, while vertical ones are still, as it is expected. Some numerical noise is noticeable, as well as slight rounding of the sharp corners. These effects can be attributed to the choice of the difference schemes used in our simulations. It can be noticed, also, that these effects are much less pronounced in our simulation, than it is refereed in [5]. For longer etching times rounding of the corners become significant and it is obvious that more appropriate difference scheme should be used.

A more realistic etching rate angular dependent simulation results are presented in Figure 4. Velocity function in the form $F(t, \mathbf{x}) = cos(\theta) [1 + 4 \, sin^2(\theta)]$ is usual for the ion milling processes. This is a typical example of non-convex Hamiltonian, where upwind difference scheme shouldn't be applied. Nevertheless, results of our simulations are quite satisfactory even in that case. The faceting in the upper part of the trench is reproduced correctly, but the small deformations of the bottom are obvious.

**CONCLUSIONS**

In this paper we have presented some preliminary calculations of the 3D etching profile evolution based on sparse-field level set method. The obtained results show correct behavior for simple etching models considered here, even in the case of non-convex Hamiltonian, when more pronounced numerical noise and deformations at the corner are observed. In such cases it would be necessary to use Lax-Friedrichs difference scheme, instead of upwind schemes used our simulations.